# Quasiclassical quantization of the Boussinesq breather emerging from the kink localized mode

M.M. Bogdan[1], O.V. Charkina[1,2]

[1]*B. Verkin Institute for Low Temperature Physics and Engineering of the National Academy of Sciences of Ukraine, 47 Nauky Ave., Kharkiv, 61103, Ukraine*

[2] *University of Luxembourg, L-1511 Luxembourg City, Luxembourg*

E-mail: charkina@ilt.kharkov.ua
bogdan@ilt.kharkov.ua

## Abstract

The breather solution found by M. Tajiri and Y. Murakami for the Boussinesq equation is studied analytically. The new parameterization of the solution is proposed, allowing us to find exactly the existence boundary of the Boussinesq breather and to show that such a nonlinear excitation emerges from the linear localized mode of the kink solution corresponding to a shock wave analog in a crystal. We explicitly find the first integrals, namely the energy and the field momentum, and faithfully construct the adiabatic invariant for the Boussinesq breather. As a result, we carry out the quasiclassical quantization of the nonlinear oscillating solution, obtaining its energy spectrum, i.e., the energy dependence on the momentum and the number of states, and reveal the Hamiltonian equations for this particle-like excitation.

## Introduction

The significant progress at the beginning of the soliton investigations a half-century ago was achieved due to the development of mathematical methods of exact solving the integrable equations. The physical theory of solitons required universal methods of analysis of the non-integrable nonlinear models and approximate solving the corresponding underlying equations. The important role in the solution of the problem in the nonlinear theory belongs to the paper by A.M. Kosevich and A.S. Kovalev [1], in which the asymptotic procedure of finding the self-localized nonlinear excitations, called later the breathers, was formulated. Nowadays, we meet the 80th anniversary of Prof. A.S. Kovalev, the favorite scholar and colleague of Prof. A.M. Kosevich, and the fiftieth anniversary of the appearance of their fundamental work. It is a great honor for us to contribute to the present anniversary issue devoted to our remarkable colleague and friend Prof. A.S. Kovalev.

The ideas of Ref. 1 gave rise to several directions in the dynamical theory of nonlinear excitations. The present work aims to demonstrate that they work well today, too. After Ref. 1 the small-amplitude breather solutions can be constructed in different continuum models of elastic and magnetic media. As a system for the demonstration of the efficiency of the asymptotic method, the authors chose the nonlinear model of a one-dimensional crystal with the following energy

$$E_{cr} = \frac{m}{2}\sum_n \left(\frac{du_n}{d\tau}\right)^2 + \sum_n U(u_n - u_{n-1}) + \sum_n U_{ext}(u_n), \tag{1}$$

where $m$ is an atomic mass, $u_n$ is its displacement of $n$-th atom, $\tau$ is time, the first term is kinetic energy, the second is the potential energy of nonlinear interaction of nearest neighboring atoms and the last term is a nonlinear external potential. The breather solution was constructed for the nonlinear Klein-Gordon equation in a crystal with Hooke's law of interaction between neighbors and a general type of small-amplitude nonlinear potential. The exact sine-Gordon breather could be considered as a special case of such a solution. The authors also studied the self-localized nonlinear vibrations with frequencies slightly above the upper edge of the continuous wave spectrum and predicted the existence of the discrete breathers [2]. At last, they showed that the small-amplitude breathers can exist in a crystal with only a nonlinear interatomic interaction in the absence of the external potential. In the crystal the potential energy can be expanded as follows [3]

$$U(u_n - u_{n-1}) = \frac{1}{2}U''(0)(u_n - u_{n-1})^2 + \frac{1}{6}U'''(0)(u_n - u_{n-1})^3 + \frac{1}{24}U''''(0)(u_n - u_{n-1})^4 \tag{2}$$

where primes denote taking the derivatives. In this case the condition of existence of the breather-like envelopes of the antiphase atomic oscillations was established, which states that only the cubic potential energy cannot produce such a type of self-localized oscillations. It needs the quartic term in the potential energy. The Lagrange equations of motion of the lattice with the potential energy in the presence of only cubic or quartic term describe the dynamics of the Fermi-Pasta-Ulam $\alpha$- or $\beta$-models, respectively [4]. In the long-wavelength limit, these Lagrange equations are reduced to the Boussinesq and modified Boussinesq equations. Typically, in the lattice theory [3], only the first anharmonic term remains in the expansion (2), then the crystal energy is written as the following

$$E_{tot} = \int_{-\infty}^{\infty} m\frac{dz}{a}\left\{\left(\frac{1}{2}u_\tau^2 + \frac{s^2}{2}u_z^2 - \frac{B^2}{2}u_{zz}^2 - \frac{\Lambda^2}{6}u_z^3\right)\right\} \tag{3}$$

where the subscripts denote the derivatives, $z = an$ is a coordinate along the atomic chain, $a$ is the lattice constant, and the second and third derivatives of the potential (2) determine the sound velocity $s$, the parameter of the higher dispersion $B$, and the coefficient of nonlinearity $\Lambda$ as follows

$$s^2 = \frac{a^2}{m}U''(0), \; B^2 = \frac{a^2}{12}s^2, \; \Lambda^2 = -\frac{a^3}{m}U'''(0). \tag{4}$$

The dimensional Boussinesq equation, having the first integral $E_{tot}$, takes the form

$$u_{\tau\tau} - s^2 u_{zz} - B^2 u_{zzzz} + \Lambda^2 u_z u_{zz} = 0 \tag{5}$$

First, Boussinesq derived his equation to describe the wave motion in shallow water. In the dimensionless variables, the equation (5) takes the form

$$\varphi_{tt} - \varphi_{xx} - \varphi_{xxxx} + 6\varphi_x \varphi_{xx} = 0 \qquad (6)$$

The relations between parameters of Eqs. (5) and (6) are given in the next section.

It appeared that the Boussinesq equation is exactly integrable [5, 6] and possesses the multisoliton solutions [7] and series of conservation laws [8], despite its "bad" dispersion law $\omega(k) = k\sqrt{1-k^2}$ for the continuous spectrum waves. The state of rest with $\varphi = 0$ is unstable with respect to linear waves with the wave number $k > 1$. Therefore, Eq. (6) is called the "ill-posed" Boussinesq equation. The one-soliton solution of Eq. (6) corresponds to the kink, which is the shock wave analog in a crystal in terms of Eq. (5). The first spatial derivative $\partial u/\partial z$ means the crystal deformation, which is negative for the shock wave. The problem of the soliton stability was solved by J. Berryman in the framework of the linearized Boussinesq equation [9]. Berryman's analysis showed that soliton is unstable and a small addition to the solution grows exponentially. However, N. Yajima studied the nonlinear stage of the instability and found the growing mode from the multisoliton formula, which at first grows in time but then quickly disappears [10].

The Boussinesq equation (6) can be reduced approximately to the integrable Korteweg-de Vries equation [11]. In its turn, the integrable discrete equation of the Toda lattice in the small-amplitude limit transforms to the Boussinesq equation if we substitute the Toda potential into Eq. (2). All the equations possess the standard multisoliton formulas for nonlinear "superposition" of the usual one-parametric solitons. The breather solutions exist in the integrable modified Korteweg-de Vries equation with the cubic nonlinear term [11]. Formally, its breather can be obtained from the two-soliton formula by the transition of soliton parameters to the complex plane, keeping the solution real. Constructing the approximate solution for the moving breather in the modified Boussinesq equation is possible using the improved asymptotic procedure [12].

The attempts to construct breathers in the Korteweg-de Vries and Toda equations led to the complexiton solutions [13–16]. This kind of solutions is characterized by the complexity of the spectral parameters in the associated inverse scattering problem and by the singularity of soliton forms. Surprisingly, the transition to the complex conjugate parameters in a soliton pair in the Boussinesq equation gave rise to the real regular oscillating solution. M. Tajiri and Y. Murakami revealed this breather solution with cumbersome relations between its parameters, verified its validity, and found numerically a boundary of existence, and showed that the Yajima growing mode did not destroy the solution [17]. To our knowledge, only recurring works by O.V. Kaptsov [18,19] contain a numerical investigation of the form of this solution for some set of its parameters.

In this contribution, we propose a new parameterization of the found breather solution, which significantly simplifies the realization of an analytical approach to research of dynamical characteristics of the breather solution. We find analytically the existence boundary of the breather, previously obtained only numerically. We show that this kind of the breathing solution is of a special type, because it is the first time when the found expression explicitly demonstrates how a nonlinear excitation emerges from the linear localized mode of the kink. We propose to call this solution the Boussinesq breather. We explicitly calculate the first integrals, the energy and the field momentum, and the adiabatic invariant for the Boussinesq breather. Then we perform the quasiclassical quantization of the breather, getting its energy spectrum, i.e. the energy dependence on the number of states and the momentum, and bringing out the Hamiltonian equations for this particle-like excitation.

**The particle-like kink of the Boussinesq equation and its Hamiltonian equation**

The analysis of the one-soliton solution of the Boussinesq equation (5) as a shock wave analog in the 1D crystal is presented in Refs. 3 and 11. In terms of atomic displacements, the soliton has the kink form. The impact on a crystal at the left infinity causes a finite displacement $u_0$ of atoms, and the kink moves in medium with the supersonic velocity $v > s$, which is a single independent parameter of the solution:

$$u_K = \frac{u_0}{1+\exp\left(\frac{z-v\tau}{l}\right)} = \frac{u_0}{2}\left[1 - \tanh\frac{z-v\tau}{2l}\right], \qquad (7)$$

$$u_0 = 12\frac{v^2-s^2}{\Lambda^2}l, \qquad l = a\cdot\left[12\left(\frac{v^2}{s^2}-1\right)\right]^{-\frac{1}{2}}. \qquad (8)$$

The analog of the shock wave with effective length $l$, constricted with the velocity increase, produces the negative localized deformation that characterizes the local compression in a crystal. Because of the negative value the deformation gives the positive contribution to the crystal energy (3) through the anharmonic term.

As pointed out above, it is possible to obtain the dimensionless Boussinesq equation (6) from Eq. (5) by the appropriate changes of variables. Introducing a new variable $\varphi$ for the dimensionless displacement and measuring the energy, time, and coordinate in the following units:

$$u = aA\varphi, \quad A = \sqrt{3}\frac{s^2}{\Lambda^2}, \quad E_0 = 6\sqrt{3}m\frac{s^6}{\Lambda^4}, \quad \tau_0 = \frac{a}{\sqrt{12}s}, \quad z_0 = s\tau_0, \qquad (9)$$

we come to Eq. (6) in its traditional mathematical form and obtain the dimensionless expression for the energy as the first integral of the equation:

$$E = \int_{-\infty}^{\infty} dx \left\{ \left( \frac{1}{2} \varphi_t^2 + \frac{1}{2} \varphi_x^2 - \frac{1}{2} \varphi_{xx}^2 - \varphi_x^3 \right) \right\}. \tag{10}$$

Here $t$ and $x$ are the dimensionless time and coordinate, respectively. Now, we rewrite the one-soliton solution and its spatial derivative by the use of the traditional free parameter $K$:

$$\varphi_K = K \left[ 1 - \tanh \frac{K}{2}(x - V_K t) \right], \qquad V_K = \sqrt{1 + K^2} \tag{11}$$

$$w_K = \frac{\partial \varphi_K}{\partial x} = -\frac{K^2}{2} \frac{1}{\cosh^2 \frac{K}{2}(x - V_K t)}. \tag{12}$$

As easily seen [11], the soliton energy can be expressed completely through the deformation variable:

$$E(K) = \int_{-\infty}^{\infty} dx \left\{ \left( \frac{1}{2}(V^2 + 1) w_K^2 - \frac{1}{2} \left( \frac{\partial w_K}{\partial x} \right)^2 - w_K^3 \right) \right\}. \tag{13}$$

After taking the simple integrals, we finally find the energy dependence on the parameter $K$:

$$E(K) = \frac{2}{3} K^3 \left( 1 + \frac{4}{5} K^2 \right). \tag{14}$$

There is another integral of the Boussinesq equation, the field momentum [3,11], which is defined as

$$P = -\int_{-\infty}^{\infty} p \, d\varphi \equiv -\int_{-\infty}^{\infty} \frac{\partial L}{\partial \varphi_t} d\varphi = -\int_{-\infty}^{\infty} \frac{\partial \varphi}{\partial t} \frac{\partial \varphi}{\partial x} dx. \tag{15}$$

The integral is easily determined and can be interpreted as a product of the effective mass and velocity of the soliton.

$$P(K) = V_K \int_{-\infty}^{\infty} w_K^2 dx = \frac{2}{3} K^3 \sqrt{1 + K^2} \equiv M_K V_K \tag{16}$$

The formulas (14) and (16) set the parametric dependence between the first integrals. Calculating the derivatives of integrals with respect to parameter $K$ we find

$$\frac{dE}{dK} = 2K^2 \left( 1 + \frac{4}{3} K^2 \right), \qquad \frac{dP}{dK} = \frac{2K^2 \left( 1 + \frac{4}{3} K^2 \right)}{\sqrt{1 + K^2}}. \tag{17}$$

The energy, considering as a function of the momentum, is the Hamiltonian function $H_K(P)$. In order to calculate the derivative of the energy with respect to the momentum, we can use Eqs. (17) and (11), and establish finally the Hamiltonian equation for the soliton as the particle-like excitation [11]:

$$\frac{\partial E}{\partial P} = \frac{dE}{dK} \frac{dK}{dP} = \sqrt{1 + K^2} = V_K, \qquad X_t = V_K = \frac{\partial H_K}{\partial P}. \tag{18}$$

Here $X$ is a coordinate of a center of the soliton moving with the constant velocity $V_K$ as a particle in mechanics.

## The Boussinesq breather and its emergence from kink localized mode

Riogo Hirota found the multisoliton formula [7] for the Boussinesq equation written for the dimensionless deformation

$$w_{tt} - w_{xx} - w_{xxxx} + 3(w^2)_{xx} = 0. \tag{19}$$

Using the Hirota transformation for the variables $\varphi$ and $w$

$$\varphi = -2\frac{\partial \ln f}{\partial x}, \qquad w = -2\frac{\partial^2}{\partial x^2} \ln f, \tag{20}$$

one can obtain the Hirota bilinear equation [7] for the function $f$ and find the one-soliton solution

$$f_K = 1 + \exp K(x - V_K t), \tag{21}$$

considered in the previous section, and the two-soliton solution in the form

$$f_{2s} = 1 + \exp\eta_1 + \exp\eta_2 + a_{12}\exp(\eta_1 + \eta_2), \tag{22}$$

with the following parameters

$$\eta_i = k_i x - \Omega_i t + \eta_i^0, \qquad \Omega_i = \pm k_i\sqrt{1 + k_i^2}, \qquad V_i = \pm\sqrt{1 + k_i^2}, \qquad i = 1,2, \tag{23}$$

$$a_{12} = -\frac{(\Omega_1 - \Omega_2)^2 - (k_1 - k_2)^2 - (k_1 - k_2)^4}{(\Omega_1 + \Omega_2)^2 - (k_1 + k_2)^2 - (k_1 + k_2)^4}. \tag{24}$$

The every soliton has one independent real parameter $k_i$ apart from an arbitrary phase $\eta_i^0$. The interaction of two solitons, described by the formulas (20) and (22), is typical for the integrable equation, and it is reduced to the soliton phase shifts after their collision.

M. Tajiri and Y. Murakami, using a well-known technique [11], chose for two solitons the pair of the complex conjugate parameters $k_1$ and $k_2$ as follows

$$k_1 = \kappa + ik, \qquad k_2 = \kappa - ik \tag{25}$$

in order to construct a new real solution. Due to the supposed validity of the relations (23) and (24) for complex parameters

$$\Omega_1 = \Omega + i\omega, \qquad \Omega_2 = \Omega - i\omega, \tag{26}$$

$$\eta_1^0 = \eta_0 = (\eta_2^0)^*, \tag{27}$$

$$(\Omega + i\omega)^2 = (\kappa + ik)^2 + (\kappa + ik)^4, \tag{28}$$

they found the ponderous relations between real and imaginary parts of parameters $k_i$ and $\Omega_i$ and coefficient $a_{12}$. Therefore, they needed to use numerical analysis for the clarification of the obtained solution, its form, and the existence boundary.

In the present contribution, we propose a new parameterization for the breather solution, introducing parameters $\alpha$ and $\beta$ as follows

$$\kappa \pm ik \equiv \sinh(\alpha \pm i\beta), \qquad V_1 = V_2^* = \cosh(\alpha + i\beta), \tag{29}$$

$$\Omega_1 = \Omega_2^* = \Omega + i\omega = \pm \sinh(\alpha + i\beta)\cosh(\alpha + i\beta) = \pm \frac{1}{2}\sinh 2(\alpha + i\beta), \tag{30}$$

$$\kappa = \sinh\alpha \cos\beta, \qquad k = \cosh\alpha \sin\beta, \tag{31}$$

$$\Omega = \pm\frac{1}{2}\sinh 2\alpha \cos 2\beta, \qquad \omega = \pm\frac{1}{2}\cosh 2\alpha \sin 2\beta, \tag{32}$$

$$a_{12} = \frac{\sin^2\beta \cdot (4\cosh^2\alpha - 1)}{\sinh^2\alpha \cdot (1 - 4\cos^2\beta)}. \tag{33}$$

As a result, we have a very compact and transparent expression for coefficient $a_{12}$, the sign of which is principal for the existence of either the breather or the complexiton solution. When $a_{12} > 0$, then by taking out of brackets the expression from the function (22), and by the substitution of the function $f_{br}$ to the deformation in Eq.(20), we find the following formula for the breather solution:

$$w_{br} = -2\frac{\partial^2}{\partial x^2}\ln f_{br}, \qquad f_{br} = \cosh(\kappa x - \Omega t) + \lambda \cos(kx - \omega t), \tag{34}$$

$$\lambda = \frac{1}{\sqrt{a_{12}}} = \frac{\sinh\alpha \cdot \sqrt{1 - 4\cos^2\beta}}{\sin\beta\sqrt{4\cosh^2\alpha - 1}}, \tag{35}$$

where by the choice $\mathrm{Re}\,\eta_0 = -\ln\sqrt{a_{12}}$ we place the solution center in the origin at the time $t_0$, which is chosen as $t_0 = 0$ by setting $\mathrm{Im}\,\eta_0$ equal to zero. Note that the exponent, taken out of brackets, does not contribute to the deformation $w(x,t)$ due to the second differentiation. When $a_{12} < 0$, then instead of the function $\cosh(\kappa x - \Omega t)$ in Eq.(34), we obtain $\sinh(\kappa x - \Omega t)$ and finally come to the type of the singular solution, named by X. Ma as the complexiton [15,16].

We present the solution (34) in the form of the algebraic sum of the pure soliton and the nonlinear breathing mode:

$$w_{br} = -2\frac{\partial^2}{\partial x^2}\ln\cosh(\kappa x - \Omega t) - 2\frac{\partial^2}{\partial x^2}\ln\left\{1 + \frac{\lambda\cos(kx - \omega t)}{\cosh(\kappa x - \Omega t)}\right\}, \tag{36}$$

$$V = \frac{\Omega}{\kappa} = \pm \cosh\alpha \cdot \frac{\cos 2\beta}{\cos\beta}, \qquad v_r = \frac{\omega}{k} = \pm \cos\beta \cdot \frac{\cosh 2\alpha}{\cosh\alpha}, \qquad (37)$$

where $V$ is the velocity of the soliton envelope and $v_r$ is the phase velocity of the carrier wave in the breathing mode. Note that the parameter $\alpha$ can change formally from zero to infinity. At the same time, the range of changing the parameter $\beta$ is finite, namely, $\beta_0 \leq \beta \leq \pi/2$, because we impose the condition of positivity of the parameter $\kappa$. The origin of the critical value $\beta_0 = \pi/3$ is explained below. As a consequence, the parameters $\Omega$ and $V$ would be negative if we chose the positive sign in Eqs. (32) and (37), and simultaneously the parameters $\omega$ and $v_r$ would be positive. If we would like to get the direction of the soliton movement the same as in the case of the solution (12), we have to choose the negative sign in these formulas. In both cases the soliton envelope and the carrier wave move in opposite directions in order to keep the breather solution valid.

As easily seen from Eq. (35), when the parameter $\lambda \ll 1$, then the second function in the right-hand side of the equation (36), is reduced to the expression

$$\psi = -2\lambda \frac{\partial^2}{\partial x^2} h(x,t), \qquad h(x,t) = \frac{\cos(kx - \omega t)}{\cosh(\kappa x - \Omega t)}. \qquad (38)$$

The function $h(x,t)$ represents the typical form of the small-amplitude breather [11]. The presentation of the solution as the algebraic sum of two localized nonlinear waves with the joint center of mass explains the structural features of the complex excitation (36), found numerically in [17–19]. The typical evolution picture of the breather solution is presented in Fig.1.

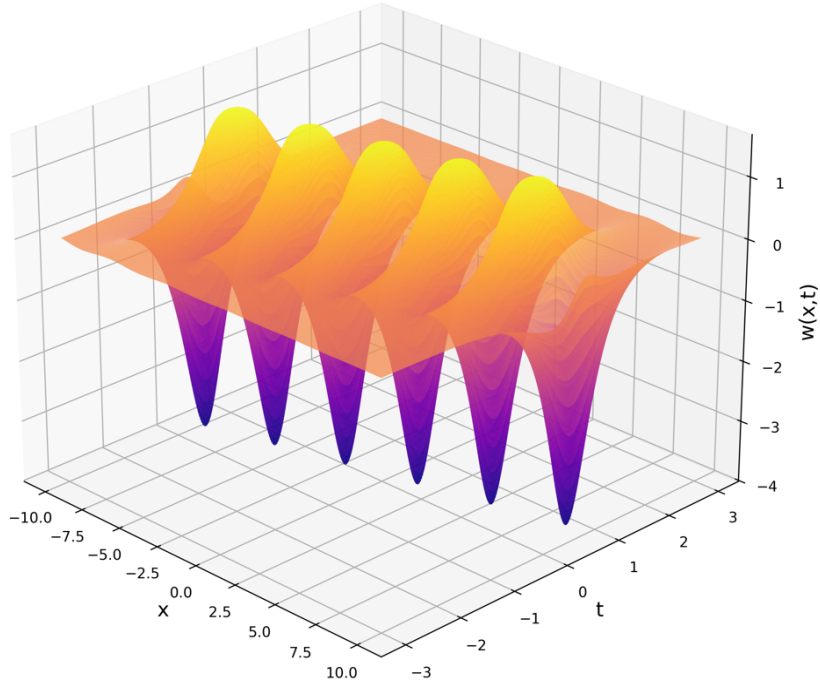

Fig. 1. The evolution of the breather solution of the Boussinesq equation (19) at the parameters $\alpha = 1.1$ and $\beta = 1.45$.

As follows from Eq. (35), the small amplitude $\lambda$ can be provided by either the smallness of the parameter $\alpha$ or by the approach of the parameter $\beta$ from the above to the critical value $\beta_0$. When $\alpha \ll 1$, then both the deformation soliton and the breather have their amplitudes of one order of the smallness. In this limit the algebraic sum of two excitations behaves like the weakly-localized wave packet, however, without dispersive spreading. When the parameter $\beta$ approaches $\beta_0$ then the breather magnitude has to vanish while the soliton can be of any arbitrary amplitude. In this limit, the superposition of the soliton and small breather looks as the caterpillar-like function. At the point $\beta_0$ the regular breather solution, defined from Eqs. (20) and (22), loses its validity and transforms to a singular one. Due to the new parameterization, now we are able to find exactly the existence boundary of the breather solution. This is the curve in the plane of the parameters $\kappa$ and $k$, corresponding to the parameter $\lambda$ equal zero due to $\beta = \beta_0$. As seen from Eq. (31) and (32), in this case the solution parameters are simplified to

$$\cos \beta_0 = \frac{1}{2}, \qquad \kappa_0 = \frac{1}{2}\sinh \alpha, \qquad k_0 = \frac{\sqrt{3}}{2}\cosh \alpha, \qquad (39)$$

$$\Omega_0 = \mp\frac{1}{2}\sinh \alpha \cdot \cosh \alpha, \qquad V_K = \frac{\Omega_0}{\kappa_0} = \mp\cosh \alpha, \qquad \omega_0 = \pm\frac{\sqrt{3}}{4}\cosh 2\alpha. \qquad (40)$$

It immediately follows from the expressions (39) that the existence boundary in terms of parameters $\kappa_0$ and $k_0$, obtained numerically in [17], is very simple:

$$k_0(\kappa_0) = \frac{\sqrt{3}}{2}\sqrt{1+4\kappa_0^2}. \tag{41}$$

At nearest vicinity of the line, we have the algebraic sum

$$w_{br} = w_K + \psi, \qquad \psi \ll 1 \tag{42}$$

of the deformation soliton

$$w_K = -\frac{2\kappa_0^2}{\cosh^2 \kappa_0(x - V_K t)}, \qquad \kappa_0 = K/2, \tag{43}$$

which is completely consistent with the expression (12), and the very small addition ($\lambda \to 0$) that is nothing but a linear excitation mode of the soliton

$$\psi_0 = -2\lambda \frac{\partial^2}{\partial x^2}\left[\frac{\cos(k_0 x - \omega_0 t)}{\cosh(\kappa_0 x - \Omega_0 t)}\right]. \tag{44}$$

In fact, if we linearize the Boussinesq equation (19) near the soliton (see Eq. (42)), then by following Berryman's theory of the instability [9], we find the linearized equation

$$\psi_{tt} - \psi_{xx} - \psi_{xxxx} + 6(w_K \psi)_{xx} = 0. \tag{45}$$

After the substitution $\psi = g_{xx}$ and the twice integration of Eq. (45) one obtains the equation

$$g_{tt} - g_{xx} - g_{xxxx} + 6w_K g_{xx} = 0. \tag{46}$$

Introducing new variables $t$ and $y = \kappa_0(x - V_K t)$ one derives the compact form of the linearized equation

$$g_{tt} - 2V_K g_{ty} + L g_{yy} = 0, \tag{47}$$

containing the well-known operator [20]:

$$L = \kappa_0^2\left(-\frac{\partial^2}{\partial y^2} + 4 - \frac{12}{\cosh^2 y}\right). \tag{48}$$

J. Berryman sought for the instability mode, exponentially growing in time, among general solutions of the equation (47). At the same time, it is easy to verify directly that the following function

$$g(y,t) = \frac{\cos(k_m y - \omega_m t)}{\cosh(y)}, \tag{49}$$

$$k_m = \frac{k_0}{\kappa_0}, \quad \omega_m = \frac{1}{\kappa_0}(\omega_0 - k_0 V_K), \tag{50}$$

is the exact solution of Eq. (47), and hence the function $\psi_0$, given by Eq. (44), is a linear localized oscillating mode of the soliton (43). The dispersion law of the carrier wave in the localized mode is found from formulas (39) and (40)

$$\omega_0(k_0) = \frac{2}{\sqrt{3}}\left(k_0^2 - \frac{3}{8}\right) \quad . \tag{51}$$

In Fig. 2 the dispersion law of the continuous spectrum waves $\omega(k)$ is indicated, as well as the curve (51) of the localized mode frequency for the equal values of the wave number $k$. As seen in Fig.2, the localized mode appears at once $\alpha > 0$ and $\kappa > 0$, and its frequency detaches from the continuous spectrum at $k_0 = \sqrt{3}/2$, then it exists as the local frequency outside the continuous spectrum. This mode is relevant to call the *external* localized mode.

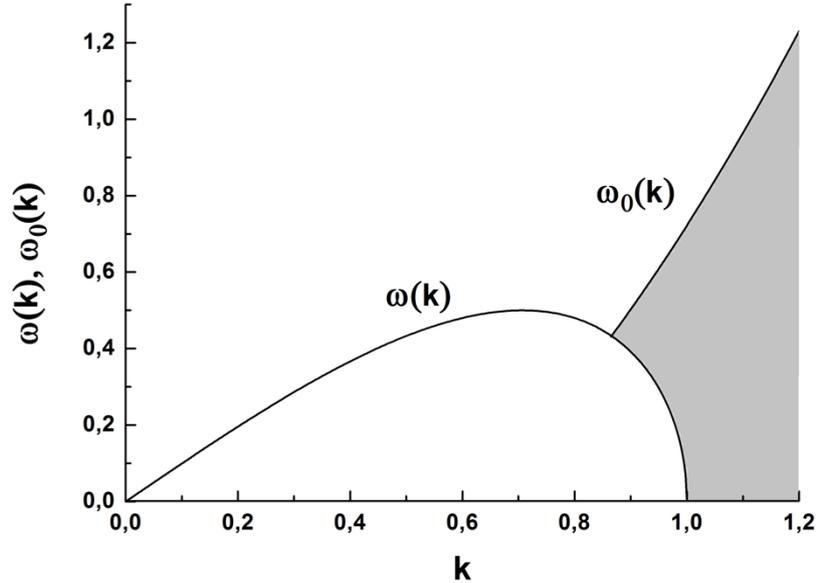

Fig.2. The dispersion law $\omega(k)$ of continuous spectrum waves and the carrier frequency $\omega_0(k)$ of the kink localized mode. The lines of these dependencies serve as boundaries of the existence area of the Boussinesq breather denoting by the grey color.

Thus, we show that this is the first explicit analytical result when the breather-like solution emerges from the linear localized mode of a soliton. For a long time the presence of the *internal* localized mode in the spectrum of linear excitations of a soliton, i.e. below the lower edge of the continuous wave spectrum, in the nonlinear models serves as a sufficient condition at least for the existence of non-trivial interaction between solitons, including the inelastic and decay processes in their dynamics. Even more, the presence the *internal* localized mode has been regarded as a sufficient condition of the non-integrability of corresponding soliton-bearing nonlinear equations [21]. For the first time, this statement was formulated in [22] and used for the investigation of the transition from the integrability to the non-integrability in the Landau–Lifshitz equation. Later, this idea was exploited in relation to the double sine-Gordon and near-discrete nonlinear Schrödinger equations [23].

However, the Boussinesq equation (19) is integrable, and despite own "ill-posedness" by reason of the "bad" linear wave spectrum, the soliton interactions in the equation are trivial, using the terminology of trivial and non-trivial interactions, adopted in V.E. Zakharov's papers [5,6]. Even the instability of the Boussinesq soliton, found by Berryman, only gave birth to the Yajima nonlinear growing mode [10], which lives a finite time and does not destroy the solution [17]. Therefore, a main unusual property of the Boussinesq equation is that its breather appears from the linear *external* localized mode of small oscillations of the soliton. In modern terms of the soliton theory, the complex breather solution (36) of the Boussinesq equation belongs to the wobbling kink family, being similar to the sine-Gordon kink-breather complex state [24–26]. In fact, returning to the displacement variable from Eqs. (20), we obtain the typical wobbling kink construction, shown in Fig. 3.

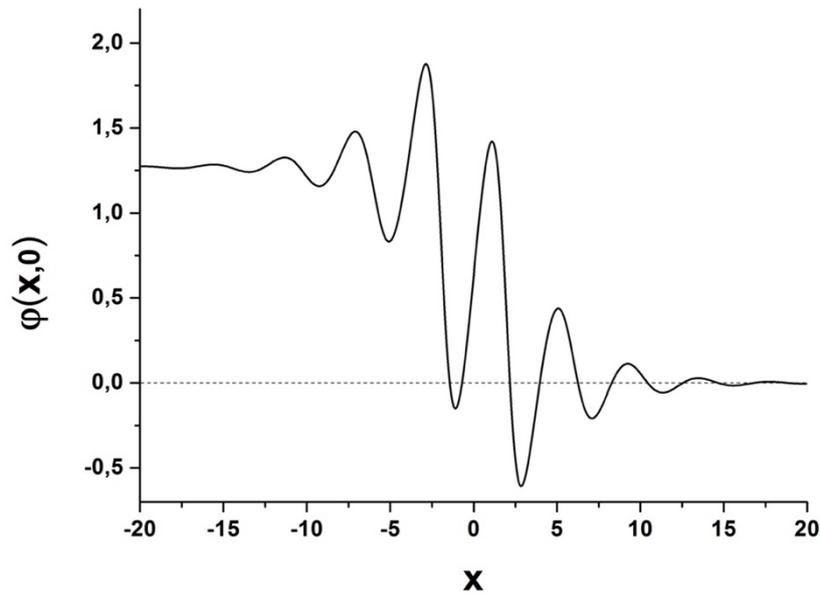

Fig. 3. The breather solution as the wobbling kink of the Boussinesq equation (6).

The breather, born from the linear *external* localized mode, cannot exist without the kink. On the other hand, the kink of the Boussinesq equation is not the topological excitation. As seen from Eq. (36), a soliton part of the solution can possess the same order of the small amplitude as the breathing part, and the Kosevich-Kovalev asymptotic procedure could be modified to catch that solution. On the other hand, the smallness of the coefficient $\lambda$ near the line (51) and the possible existence of two independent small parameters in the solution (36) prompts the ways of asymptotic construction of analogous solutions from the *internal* localized modes of topological kinks in non-integrable equations. The additive structure of the solution (36) points out also on the possibility of the dissociation processes of such bound states in the near-integrable highly-dispersive nonlinear media [21,27] and, as a result, on the allowability of the independent movement of the created high-frequency breather. All the above-mentioned specific features put the solution (36) in a separate category of nonlinear localized breathing excitations. Therefore, it would be natural to give it a special name, e.g., the Boussinesq breather.

**Integrals of motion and the quasiclassical spectrum of the Boussinesq breather**

The dynamical characteristics of the breathers, especially the energy and the field momentum, are very important for the understanding and interpretation [11,28] of these nonlinear excitations. In the present section, we calculate the energy, the field momentum, and the adiabatic invariant for the Boussinesq breather.

The integrability gives rise to an infinite set of the conservation laws. Usually, the first integrals of motion for the multisoliton excitations in the integrable nonlinear equations are expressed quite simple through the soliton parameters. The direct calculation of integrals of motion by the substitution of the multisoliton formula to the integral expressions is a very cumbersome and tiring task. However, the integrals are constants, and the calculation result does not depend on the time moment chosen in the multisoliton formula. Therefore the moment when all solitons, having obviously different velocities, are separated by infinite distances, is quite well for the calculation. This easily proves the additivity of the integrals of motion in the integrable equations. In particular, the values of the total energy and the field momentum are the sums of the corresponding values of the separate solitons.

The two-soliton solution of the Boussinesq equation, for example, possesses the total energy as follows

$$E_{12} = E(K_1) + E(K_2), \tag{52}$$

where $E(K)$ is given by the one-soliton energy expression (14). The two-soliton field moment is equal to the following one:

$$P_{12} = P(K_1) + P(K_2), \tag{53}$$

where $P(K)$ is given by the one-soliton momentum expression (16).

The transition to the complex parameters in the two-soliton solution led to the Boussinesq breather solution. It turns out that the answer on the question, what happens with the expressions for the energy (52) and the momentum (53) after the same transition, solves the problem of finding the first integrals for the Boussinesq breather. After the substitution of the parameter $k_1$ and its complex conjugate $k_2$ from Eq. (25) to the formula (52) instead of $K_1$ and $K_2$, we find the following expression for the energy of the Boussinesq breather:

$$E_{br} = E(\kappa + ik) + E(\kappa - ik), \tag{54}$$

$$E_{br} = E(\kappa, k) = \frac{4}{3}\kappa^3\left(1 + \frac{4}{5}\kappa^2\right) - 4\kappa k^2\left[1 + \frac{4}{3}(2\kappa^2 - k^2)\right]. \tag{55}$$

The validity of Eq. (55) is easily verified by the comparison of the expression with the result of the numerical calculation of the integral by the mathematical tools in the formula (10) after the substitution there the solution (34) for any values of its parameters from Eqs. (31) – (33).

Discussing the dependence of the energy (55) on the parameters $\kappa$ and $k$, it is interesting to note that in the small-amplitude limit the breather of the modified Korteweg-de Vries equation has the solution structure identical to function $h(x,t)$ from Eq. (38) [11]. Its energy dependence on the parameters $\kappa$ and $k$ is the following [29]:

$$E_{mKdV}(\kappa, k) = \frac{2}{3}\kappa^3 - 2\kappa k^2. \tag{56}$$

The expression (55) naturally has a more complicated form than Eq. (56), but they are practically identical in the main approximation up to a constant. This fact demonstrates the general dynamical properties of the equations with the higher dispersion and the closeness of mechanisms of the formation of the breather states in spite of the different character of the quadratic and cubic nonlinear terms in the Boussinesq equation and the modified Korteweg-de Vries equations, respectively. At the same time, we emphasize that the mechanism of the breather formation in the modified Korteweg-de Vries equation was studied well enough [30], while the mechanism of emergence of the Boussinesq breather from the external localized mode of the soliton in the method of the inverse scattering transform remains evidently undiscussed.

The energy $E_{br}$ can be written also through parameters $\alpha$ and $\beta$, for example, as follows

$$E_{br} = E(\alpha, \beta) = \frac{4}{15}\sinh\alpha \cos\beta \cdot \left[5\cosh^2\alpha \cdot (4\cosh^2\alpha - 3) - \\ - (\cos^2\beta + \sin^2 2\beta) \cdot (16\cosh^4\alpha - 12\cosh^2\alpha + 1)\right]. \tag{57}$$

The representation (57) seems to be more complex than Eq. (55), however it turns out to be productive in the derivation of the Hamiltonian equations for the Boussinesq breather. The dependence of the breather energy on the parameters $\alpha$ and $\beta$ is shown in Fig. 4.

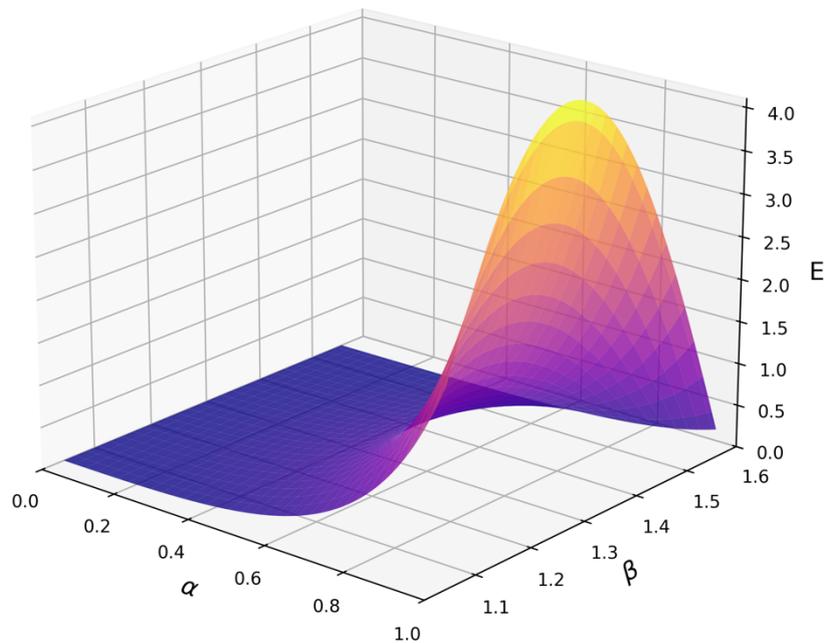

Fig. 4. The Boussinesq breather energy as a function of the parameters $\alpha$ and $\beta$.

The field momentum of the breather can be calculated directly after the substitution of the solution (34) to its integral expression (15). On the other hand, we can use the formula (53) for the two-soliton momentum to obtain the Boussinesq breather momentum by performing the transition to the complex conjugate parameters $k_1$ and $k_2$ given by Eq. (25)

$$P_{br} = P(\kappa + ik) + P(\kappa - ik). \tag{58}$$

The final expression for this integral of motion is the following:

$$P_{br} = P(\alpha, \beta) = \frac{1}{3}\sinh 2\alpha (\cosh 2\alpha \cos 4\beta - \cos 2\beta). \tag{59}$$

Again, the validity of expression (59) is confirmed by the numerical integration in the formula (15) after the substitution there the solution (34) for any values of its parameters from Eqs. (31) – (33), and successful comparing the results. The dependence of the Boussinesq breather momentum on the parameters $\alpha$ and $\beta$ is shown in Fig.5.

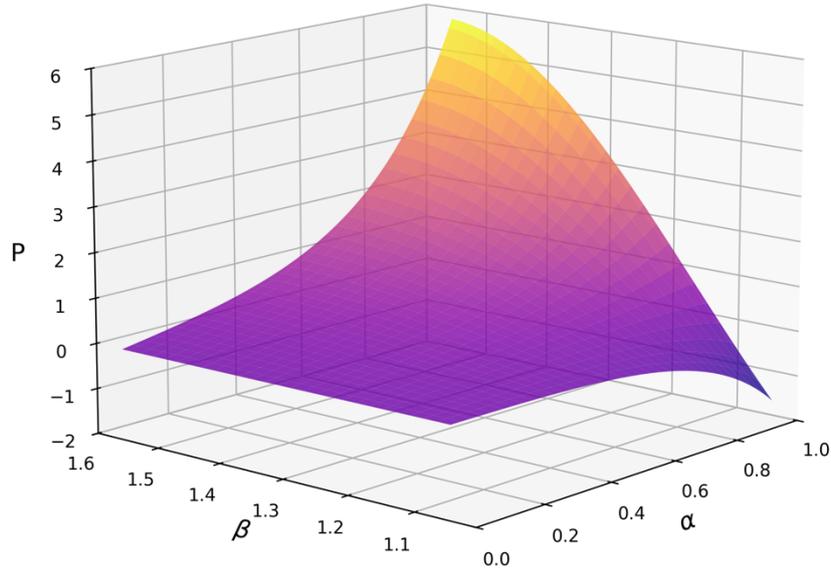

Fig. 5. The Boussinesq breather momentum as a function of the parameters $\alpha$ and $\beta$.

Now, we calculate one more important dynamical characteristic for the Boussinesq breather, namely, its adiabatic invariant. In the case of the moving oscillating complex soliton like the Boussinesq breather, the adiabatic invariant is calculated in the reference frame moving with the velocity $V$ of the envelope of this soliton. In order to construct the adiabatic invariant, we take the expression for momentum density, corresponding to the Boussinesq breather solution $\varphi_{br}(x,t)$ from Eqs. (20) and (34) as follows

$$p_{br}(x,t) = \frac{\partial \varphi_{br}(x,t)}{\partial t} \tag{60}$$

and rewrite the density $p_{br}(x,t)$ as $p_{br}(\xi,t)$, introducing the new coordinate $\xi = x - Vt$ and remaining the time equal to the previous one: $t_{new} = t$. Using the definition of the adiabatic invariant as the truncated action [11, 28], we obtain the following expression for the adiabatic invariant of the Boussinesq breather:

$$I_{br} = \int_{-\infty}^{\infty} J_{br}(\xi)d\xi = \int_{-\infty}^{\infty}\left(\frac{1}{2\pi}\oint p_{br}d\varphi_{br}\right)d\xi ,\qquad(61)$$

$$J_{br}(\xi) = \frac{1}{2\pi}\int_0^{2\pi} p_{br}(\xi,t)\frac{\partial \varphi_{br}}{\partial \theta}d\theta = \frac{1}{2\pi\tilde{\omega}}\int_0^{2\pi} p_{br}(\xi,t)\cdot\frac{\partial \varphi_{br}(\xi,t)}{\partial t}d\theta ,\qquad(62)$$

where the phase $\theta = \tilde{\omega}t$ and $\tilde{\omega}$ is the frequency of the purely periodic oscillation in the moving reference frame:

$$\tilde{\omega} = \omega - kV = \tan\beta \cdot \left(\cosh^2\alpha - \cos^2\beta\right).\qquad(63)$$

After finding the derivatives and the substitution them in Eqs. (60) and (62) we have calculated directly integrals and found the adiabatic invariant of the Boussinesq breather in the following form:

$$I_{br} = I(\alpha,\beta) = \frac{16}{9}\sinh\alpha\sin\beta\cdot\left(4\cosh^2\alpha - 1\right)\cdot\left(\sin^2\beta - \frac{3}{4}\right).\qquad(64)$$

As before, the checking of the expression (64) is fulfilled by using the Maple program, performing the numerical differentiation and integration in the formula (60) and (62) after the substitution there the solution from formulas (20), (34) and (35) for any values of its parameters from Eqs. (31) – (33). The dependence of the adiabatic invariant for the Boussinesq breather on the parameters $\alpha$ and $\beta$ is shown in Fig.6.

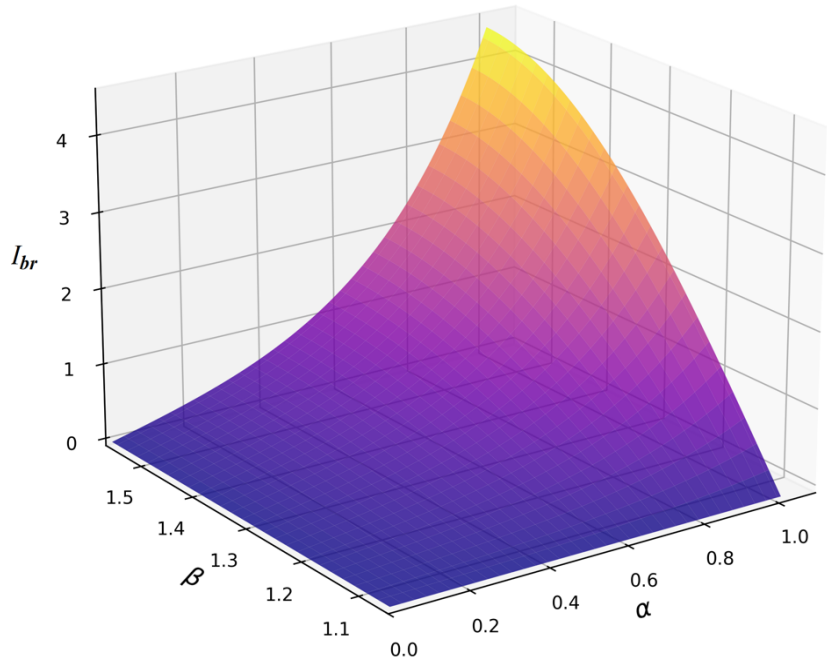

Fig. 6. The adiabatic invariant for the Boussinesq breather as a function of the parameters $\alpha$ and $\beta$

Now we are able to perform the quasiclassical quantization of the Boussinesq breather. The quasiclassical quantization of the small-amplitude self-localized vibrations in a one-dimensional atomic chain placed in the anharmonic external potential was carried out in [31]. In the framework of the integrable models, applicable to the crystal dynamics, the quasiclassical quantization was performed for the continuum and discrete breathers in the modified Korteweg-de Vries equation and the Hirota lattice equation, respectively [29,32]. In all the above cases the procedure of the quantization was fulfilled in accordance with the Bohr–Sommerfeld condition [20]. It is clear that the dimensional adiabatic invariant $I_d$ for the breather is measured in units $I_0 = E_0 \tau_0$ as seen from Eqs. (9). The Bohr–Sommerfeld quantization rule requires the following relation

$$I_d = I_0 I_{br} = \hbar \mathrm{N} , \qquad \mathrm{N} \gg 1, \tag{65}$$

where the large integer $\mathrm{N}$ denotes the number of the quantum states in the corresponding phase space of the Boussinesq breather. It is clear that due to the smallness of the dimensionless parameter $\gamma = \hbar/I_0$ the dimensionless adiabatic invariant $I_{br}$ can be of order of unity or less and considered as a quasi-continuous quantity $N$ characterizing the number of quantum states as follows

$$I_{br} = N \equiv \gamma \cdot \mathrm{N} \tag{66}$$

The Boussinesq breather energy $E = E(N,P)$ as the function of the field momentum $P$ and the normalized adiabatic invariant $N$ represents the quasiclassical energy spectrum of this complex nonlinear excitation. The quasiclassical energy spectrum $E = E(N,P)$ can be built up graphically by the use of the dependencies for the main physical quantities (57), (59) and (64) on the joint variables $\alpha$ and $\beta$. The parametrically defined energy surface, having the quite flat shape, is shown in Fig. 7.

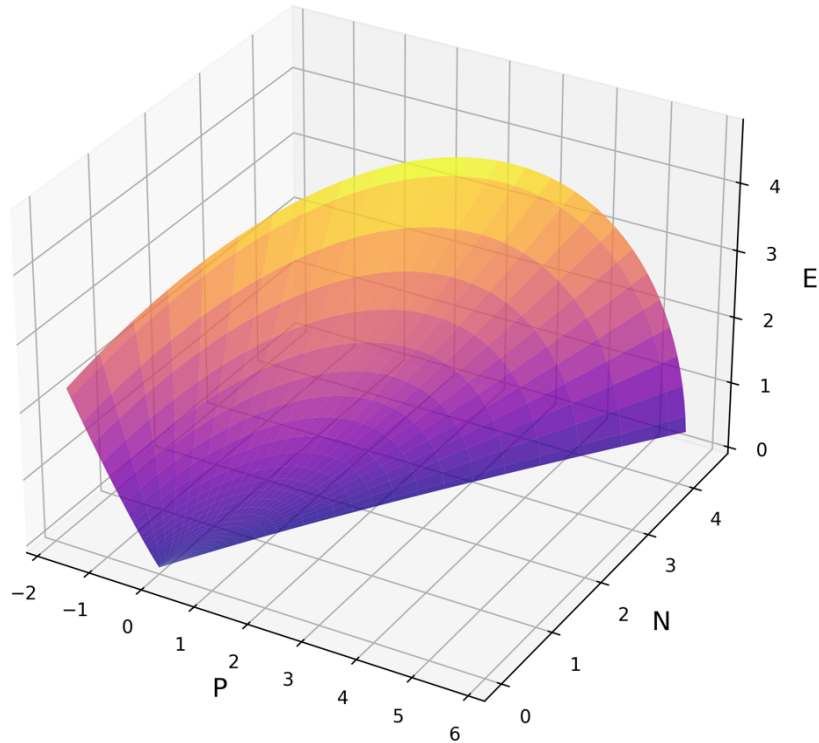

Fig. 7. The quasiclassical energy spectrum of the Boussinesq breather.

The nonlinear excitations, solitons and breathers, in the integrable models behave as the particle-like objects. The moving one-parametric soliton has a single collective coordinate $X = V_K t$ for the center of mass and the Hamiltonian equation in the form of the found equation (18). The breather possesses the additional collective variable, namely the phase $\theta$ from the equation (62).

Now, we show that such Hamiltonian equations hold for the Boussinesq breather. For simplicity, we omit the breather subscript in denotations of the energy and the momentum in the writings below. Using the explicit formulas (57), (59), (64) and (66), we easily find the partial derivatives $\frac{\partial N}{\partial \alpha}$, $\frac{\partial P}{\partial \alpha}$, and $\frac{\partial E}{\partial \alpha}$, and make sure of the validity of the relation

$$\frac{\partial E}{\partial \alpha} = \tilde{\omega} \frac{\partial N}{\partial \alpha} + V \frac{\partial P}{\partial \alpha}. \tag{67}$$

Calculating in a similar way the partial derivatives $\frac{\partial N}{\partial \beta}$, $\frac{\partial P}{\partial \beta}$, and $\frac{\partial E}{\partial \beta}$, we make sure of the validity of the following relation

$$\frac{\partial E}{\partial \beta} = \tilde{\omega} \frac{\partial N}{\partial \beta} + V \frac{\partial P}{\partial \beta}. \tag{68}$$

At last, constructing the combinations of the derivatives and variations

$$\frac{\partial E}{\partial \alpha} \delta\alpha + \frac{\partial E}{\partial \beta} \delta\beta = \tilde{\omega}\left(\frac{\partial N}{\partial \alpha} \delta\alpha + \frac{\partial N}{\partial \beta} \delta\beta\right) + V\left(\frac{\partial P}{\partial \alpha} \delta\alpha + \frac{\partial P}{\partial \beta} \delta\beta\right) \tag{69}$$

we are convinced in the identity of the variations in the left and the right sides of Eq. (69), i.e. in the validity of the following equation

$$\delta E = \tilde{\omega} \delta N + V \delta P. \tag{70}$$

Finally, keeping in mind that the Hamiltonian function $H(N,P) = E(N,P)$, we find directly from Eq. (70) the Hamiltonian equations for the Boussinesq breather as a particle-like excitation:

$$\frac{dX_{br}}{dt} = V = \frac{\partial H}{\partial P}, \qquad \frac{d\theta_{br}}{dt} = \tilde{\omega} = \omega - kV = \frac{\partial H}{\partial N}. \tag{71}$$

More than once the Hamiltonian equations for the solitons and breathers were used as a basis for qualitative description of the dynamics in the near-integrable nonlinear equations [33].

## Conclusion

The main findings of this study are as follows:
1. The analytical investigation of the dynamical properties of the breather solution found by M. Tajiri and Y. Murakami for the Boussinesq equation has been carried out in detail. This became possible due to proposing the new parameterization of the solution, which allowed us to show the composite structure of the solution, to find exactly its existence boundary, and to calculate explicitly all the dynamical characteristics of the complex excitation.

2. It turns out that the complex solution represents analytically the algebraic sum of the kink and the breather expressions. The variability of the solution forms, previously revealed in the numerical simulation, is simply explained by the difference in the amplitudes of these components. In modern terms of the soliton theory this complex solution would be related to the category of the wobbling kinks. However, in the Boussinesq equation the breather part of the solution cannot exist separately from the kink. Taking into account also its other unusual dynamical properties, we single out the complex oscillating solution into a special category and call it the Boussinesq breather.

3. We have found exactly the existence boundary of the Boussinesq breather on the plane of the dimensionless parameters of the kink inverse length and the carrier wave number of the breather. We show that at the nearest vicinity of the existence boundary the breather emerges from the *linear localized mode* of the single soliton, and this is the first time when such a birth process is demonstrated analytically. This fact encourages the search for similar exact solutions in other integrable equations, first of all, with higher dispersive terms.

4. The revealed presence of the new independent small parameter in the Boussinesq breather near the existence boundary prompts the ways of asymptotic construction of analogous solutions from the *internal* localized modes of topological kinks in non-integrable equations. In contrast to the *external localized mode* of the Boussinesq kink with the local frequency lying above the continuous wave spectrum, the *internal* modes of topological kinks usually have frequencies below the lower edge of the continuous spectrum. The Kosevich-Kovalev asymptotic procedure could be modified to catch that new type solution keeping in mind the possibility of the existence of two independent small parameters in the theory. The evident additive structure of the Boussinesq breather probably could explain in perspective the dissociation processes of similar bound states in the near-integrable highly-dispersive nonlinear media and the independent movement of the detached high-frequency breather-like excitations.

5. We have exactly obtained the first integrals, the energy and the field momentum, for the Boussinesq breather and explicitly calculated the adiabatic invariant for the complex excitation. Using the found integral characteristics of the Boussinesq breather dynamics, we have carried out the quasiclassical quantization of the nonlinear oscillating solution, obtaining its energy spectrum, i.e., the energy dependence on the momentum and the number of states, and established the Hamiltonian equations for this particle-like excitation.

# Acknowledgments

This research was funded in part, by the Luxembourg National Research Fund (FNR), grant reference U-AGR-7181-00-C called INTER/MOBILITY/22/17568826/ELEMENT. For the purpose of open access, the author has applied a Creative Commons Attribution 4.0 International (CC BY 4.0) license to any Author Accepted Manuscript version arising from this submission.